\documentclass[aps,pra,floatfix,tightenlines,amsmath,amssymb,twocolumn,showpacs]{revtex4}

\usepackage{graphicx}
\usepackage{amssymb}
\usepackage{color}
\usepackage{psfrag}
\usepackage{ifsym}
\usepackage{hyperref}

\definecolor{pink}{rgb}{1,0.078,0.57}

\newcommand{\ket}[2] {| #1 \rangle_{#2}}
\newcommand{\bra}[2] {\langle #1 |_{#2}}

\newcommand{\ee}[1] {\mathrm{e}^{#1}}

\newcommand{\dg}{^{\dagger}}

\begin{document}

\title{Fourier Transform Quantum State Tomography}

\author{Mohammadreza Mohammadi$^1$, Agata M. Bra\'nczyk$^1$ and Daniel F. V. James$^1$}
\affiliation{$^1$CQIQC and IOS, Department of Physics, University of Toronto, 60 Saint George St., Toronto, Ontario M5S 1A7, 
Canada}
\email{branczyk@physics.utoronto.ca}

\date{\today}

\begin{abstract}
We propose a technique for performing quantum state tomography of photonic polarization-encoded  multi-qubit states. Our method uses a single \emph{rotating} wave plate, a polarizing beam splitter and two photon-counting detectors per photon mode. As the wave plate rotates, the photon counters measure a pseudo-continuous signal  which is then Fourier transformed. The density matrix of the state is reconstructed using the relationship between the Fourier coefficients of the signal and the Stokes' parameters that represent the state. The experimental complexity, i.e. different wave plate rotation frequencies, scales linearly with the number of qubits.
\end{abstract}


\maketitle

Quantum state preparation is an essential ingredient in the realization of quantum technologies such as quantum computing \cite{Kok2010}, quantum cryptography \cite{Bennett1984} and other quantum information protocols \cite{Nielsen2000}. A crucial aspect of reliable state preparation is the ability to accurately characterize the state of a quantum system. To this end, quantum state tomography (QST) allows the reconstruction of a state's density matrix from measurement statistics accumulated through repeated independent measurements of multiple identically-prepared systems \cite{Vogel1989,Leonhardt1995,James2001}. 

In linear-optics, where quantum information is encoded in the polarization of a single photon, different measurement settings are realized with a combination of linear optical elements such as wave plates, beam splitters and polarizing beam splitters, followed by photon counting. QST was first accomplished in such systems by White \emph{et al.}  \cite{White1999}, where the measurement settings corresponded directly to the Stokes' parameters used to characterize the polarization state of the classical electromagnetic field \cite{Stokes1852}. Later it was suggested that an over-complete symmetric six-measurement set \cite{deBurgh2008} or an informationally-complete symmetric four-measurement set \cite{Rehacek2004,deBurgh2008,Ling2006,Medendorp2011,Kalev2012a} be used for improved performance. Other extensions, such as those considering optimal experimental design under realistic technical constraints \cite{Kosut2004,Nunn2010}, or modifications due to inaccessible information \cite{Adamson2007,Adamson2008,Asorey2011,Branczyk2012a,Mogilevtsev2012,Teo2012,Merkel2012} or preferable measurements choices \cite{deBurgh2008,Ling2006,Medendorp2011,Altepeter2005,Rehacek2007,Adamson2010,Bogdanov2010, Yamagata2011, Brida2011,Bogdanov2011,Kalev2012} have also been considered.

To date, all implementations of QST of photonic polarization-encoded qubits have utilized either multiple wave-plates and/or multiple beam splitters per qubit. We propose a technique that uses only one wave plate and one polarizing beam splitter (PBS) per qubit mode. Each mode $m$ is incident on a single wave plate rotating at frequency $\Omega_{m}$ followed by a polarizing beam splitter (PBS). Photon counters at the output ports of the PBS measure a pseudo-continuous signal and the state is reconstructed from the Fourier coefficients of this signal. The experimental complexity of this method scales linearly with the number of qubits in terms of the number of settings required (i.e. wave plate rotation frequencies) rather than exponentially, as is the case with QST that uses discrete measurement settings. Similar techniques that rely on rotating wave-plates are used in classical optics to determine the polarization state of the electromagnetic field \cite{Flueraru2008}. In the context of non-classical light, Fourier spectroscopy has been used to characterize the joint spectrum of photons \cite{Wasilewski2006a}.

The remainder of this paper is organised as follows. In Section \ref{sec:QuantumStateTomography}, we give a brief review of QST of multi-qubit states. In Section \ref{sec:FourierTransformTomography}, we introducing our scheme for Fourier transform tomography (FTT)  and provide examples for one and two qubits. In Section \ref{sec:Conclusion}, we provide concluding remarks 

\section{Quantum State Tomography}\label{sec:QuantumStateTomography}

Tomography is the process of constructing a representation of an object by imaging it in different sections. In quantum state tomography, we aim to construct a representation of a quantum state $\hat\rho$ from different measurement outcomes. An $n$-qubit system is specified by $4^n-1$ real parameters. We therefore require at least this many outcomes of linearly independent measurements to specify $\hat\rho$.

\begin{figure}[t]
\includegraphics[width=0.9\columnwidth]{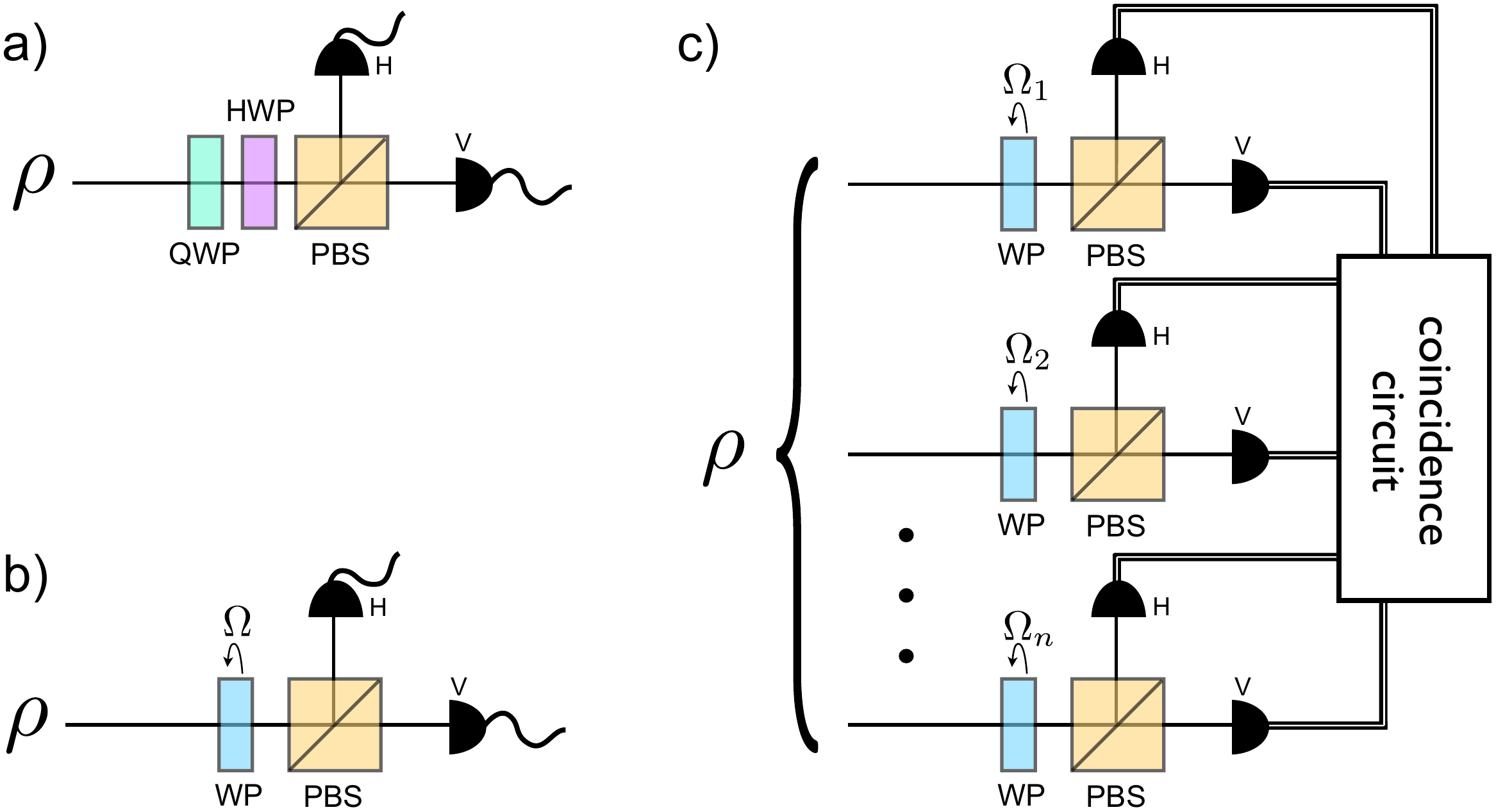}
\caption{ (Color online) Schematic diagrams of: a) typical QST set-up which uses a combination of quarter- and half-wave plates to perform arbitrary-basis measurements; b) FTT set-up which uses one rotating wave plate; c) multi-qubit FTT which uses one wave-plate per qubit mode $m$, rotating at frequency $\Omega_{m}$.  }
\label{fig:Fig1}
\end{figure}

The probability of obtaining measurement outcome $j$, given a measurement operator $\hat {M}_{j}$, is given by
\begin{eqnarray}\label{eq:ProbabilityGeneral}
{p}_{j}=\langle \hat {M}_{j}\rangle= \mathrm{Tr}[\hat \rho \hat {M}_{j}]=\frac{{n}_{j}}{\mathcal {N}_{j}}\,,
\end{eqnarray} 
where ${n}_{j}$ is the number of counts and $\mathcal {N}_{j}$ is a constant dependent on the detector efficiency and duration of data collection. In a polarization-encoded linear optical system, any projective measurement can be realized with a quarter-wave plate, a half-wave plate and a polarizing beam splitter, as shown in FIG. \ref{fig:Fig1} a). A popular choice corresponds to the three Pauli operators. 

We can always write the density matrix of an $n$-qubit system in terms of Hermitian operators $\hat\sigma_i$
\begin{eqnarray} \label{eq:RhoGeneral}
\hat \rho=\frac{1}{2^{n}}\sum_{{i}_{1},\dots, {i}_{n}={0}}^{3} S_{{i}_{1},\dots,{i}_{n}} \hat \sigma _{{i}_{1}}\otimes \dots \otimes\hat \sigma _{{i}_{n}}\,,  
\end{eqnarray}
where $\hat\sigma_{0}=\ket{H}{}\bra{H}{}+\ket{V}{}\bra{V}{}$ is the identity operator and $\hat \sigma_{1-3}$ are the Pauli operators: $\hat\sigma_{1}=\ket{H}{}\bra{V}{}+\ket{V}{}\bra{H}{}$,  $\hat\sigma_{2}=i(\ket{V}{}\bra{H}{}-\ket{H}{}\bra{V}{})$ and  $\hat\sigma_{3}=\ket{H}{}\bra{H}{}-\ket{V}{}\bra{V}{}$. The coefficients $S_{i_{1}, ...,i_{n}}=\mathrm{Tr}[\hat \rho(  \hat \sigma _{{i}_{1}}\otimes \dots \otimes\hat \sigma _{{i}_{n}})]$ completely characterize the state. $S_{i_{1}, ...,i_{n}}$ are normalized generalizations of the classical parameters introduced by Stokes  in 1852 \cite{Stokes1852}, and will hereafter be simply referred to as \emph{Stokes' parameters}. 

Combining Equations   (\ref{eq:ProbabilityGeneral}) and (\ref{eq:RhoGeneral}), we find a linear relationship between Stokes' parameters and the probability ${p}_{j}$:
\begin{align}\label{eq:ProbabilityStokes}
\begin{split}
{p}_{j}={}&\frac{1}{2^{n}}\sum_{{i}_{1},\dots, {i}_{n}={0}}^{3} S_{{i}_{1},\dots,{i}_{n}}\mathrm{Tr}[ \hat \sigma _{{i}_{1}}\otimes \dots \otimes\hat \sigma _{{i}_{n}} \hat {M}_{j}]\,,
\end{split}
\end{align}
where $\hat {M}_{j}$ acts on the entire multi-qubit system. By making $4^{n}{-}1$ linearly independent measurements, it is possible to solve for Stokes' parameters and reconstruct the density matrix according to Equation (\ref{eq:RhoGeneral}). This can be achieved through a variety of methods, including simple linear inversion, least-squares estimation or the popular maximum likelihood estimation method \cite{Hradil1997}. Alternatively, one can look to a growing number of exciting new techniques such as the forced purity routine \cite{Kaznady2009}, Baysean mean estimation \cite{Blume-Kohout2010}, compressed sensing \cite{Gross2010a}, von Neumann entropy maximization \cite{Teo2011}, hedged maximum likelihood estimation \cite{Blume-Kohout2010a}, minimax estimation \cite{Ng2012}, and techniques that focus on reconstructing the state with reliable error bars \cite{Christandl2011} and confidence regions \cite{Blume-Kohout2012}. 

\section{Fourier Transform Tomography}\label{sec:FourierTransformTomography}

In this section, we show how the quantum state of a multi-qubit system can be represented by a single joint-probability signal and how the measurement of this signal enables the reconstruction of the quantum state.

In our proposal, identical copies of the state are prepared and subsequently pass through a series of optical elements. For a multi-photon state, each photon mode $m$ is incident on a single wave plate rotating at frequency $\Omega_{m}$ followed by a polarizing beam splitter (PBS). Photon counters at the output ports of the PBS continuously measure the intensity, which can be processed to recover Stokes' parameters. A schematic of this setup is shown in in FIG \ref{fig:Fig1} b) for a single qubit and \ref{fig:Fig1} c) for multiple qubits. For multiple qubits, the signal measured is a ``coincidence intensity'' corresponding to the joint probability of detecting photons at each PBS.

\begin{figure}[t]
\includegraphics[width=0.7\columnwidth]{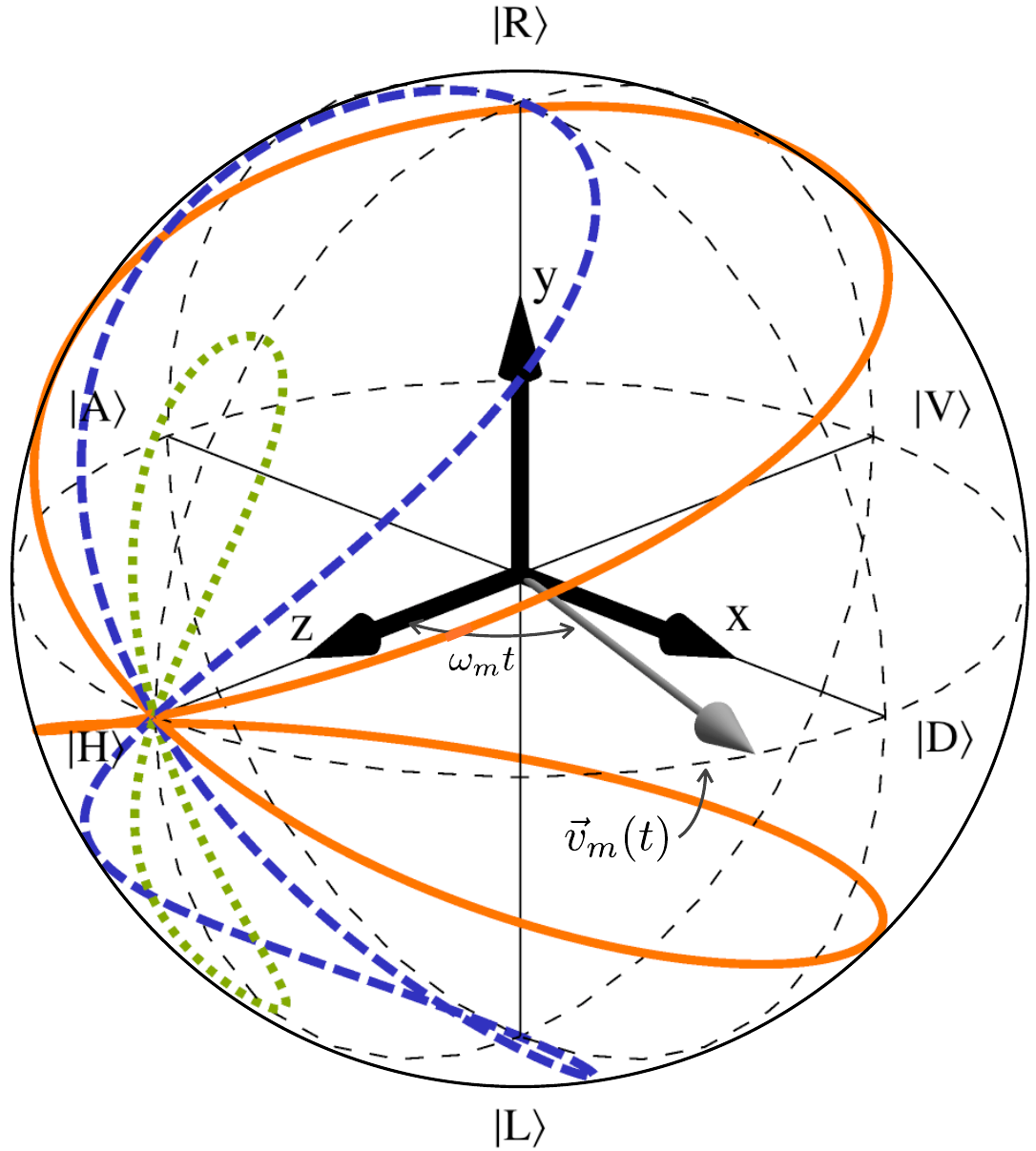}
\caption{(Color online) Path traced out by $\hat{M}^{(H)}_{m}(t)$, defined in Equation (\ref{eq:MeasurementOperator}), for: $\beta=\pi/4$ (green, dotted); $\beta=\pi/2$ (blue, dashed); and $\beta=11\pi/15$ (orange, solid).   }
\label{fig:Fig2}
\end{figure}

The time-dependent single-qubit projection-valued measure (PVM) associated with the probability of detecting a photon in the horizontal or vertical output modes of each PBS is given by $\{\hat{M}_{m}^{H}(t),\hat{M}_{m}^{V}(t)\}$ where
\begin{align}\label{eq:MeasurementOperator}
\hat{M}^{a}_{m}(t)={}&\hat{U}\dg_{m}\ket{a}{}\bra{a}{}\hat{U}_{m}\,,
\end{align}
for ${a}=H,V$, where $m$ labels the qubit mode and
\begin{eqnarray}\label{eq:Unitary}
 \hat U_{m}(t)= \cos \left (\frac {\beta}{2}\right )\hat \sigma_{0} -i\sin \left (\frac{\beta}{2}\right )\vec {v}_{m}(t) \cdot \vec{\sigma}  
\end{eqnarray}
is the unitary operator associated with a wave plate in mode $m$.  $\hat U_{m}(t)$ rotates the operators $\ket{a}{}\bra{a}{}$ on the Bloch sphere by an angle $\beta$, about the vector 
\begin{align}
\vec {v}_{m}(t)=\cos( \omega_{m}{t}) \vec k+\sin(\omega_{m}{t}) \vec i\,,
\end{align}
where $\vec k$ and $\vec i$ are unit vectors in Euclidian space (defined by the axes in FIG \ref{fig:Fig2}) and $\vec {v} \cdot \vec{\sigma} = {v}_{1}\hat \sigma_{1}+{v}_{2}\hat \sigma_{2}+{v}_{3}\hat \sigma_{3}$. As the wave plate rotates about the beam-axis at frequency $\Omega_{m}$ in real space,  $\vec {v}_{m} (t)$ rotates about the $y$-axis in Euclidian space at frequency $\omega_{m}=2\Omega_{m}$.  We assume that the fast axis of the wave plate is aligned at 0 degrees to the horizontal as defined by the polarization of the photons. A phase factor can be included in $\vec{v}_{m}(t)$ to account for different initial alignment of the wave plate.  The resulting projector $\hat{M}^{H}_{m}(t)$ traces out a figure-8 path on the Bloch sphere, as shown in FIG \ref{fig:Fig2}. The retardance of the wave plate determines the size of the figure-8.

\begin{figure}[b]
\includegraphics[width=0.8\columnwidth]{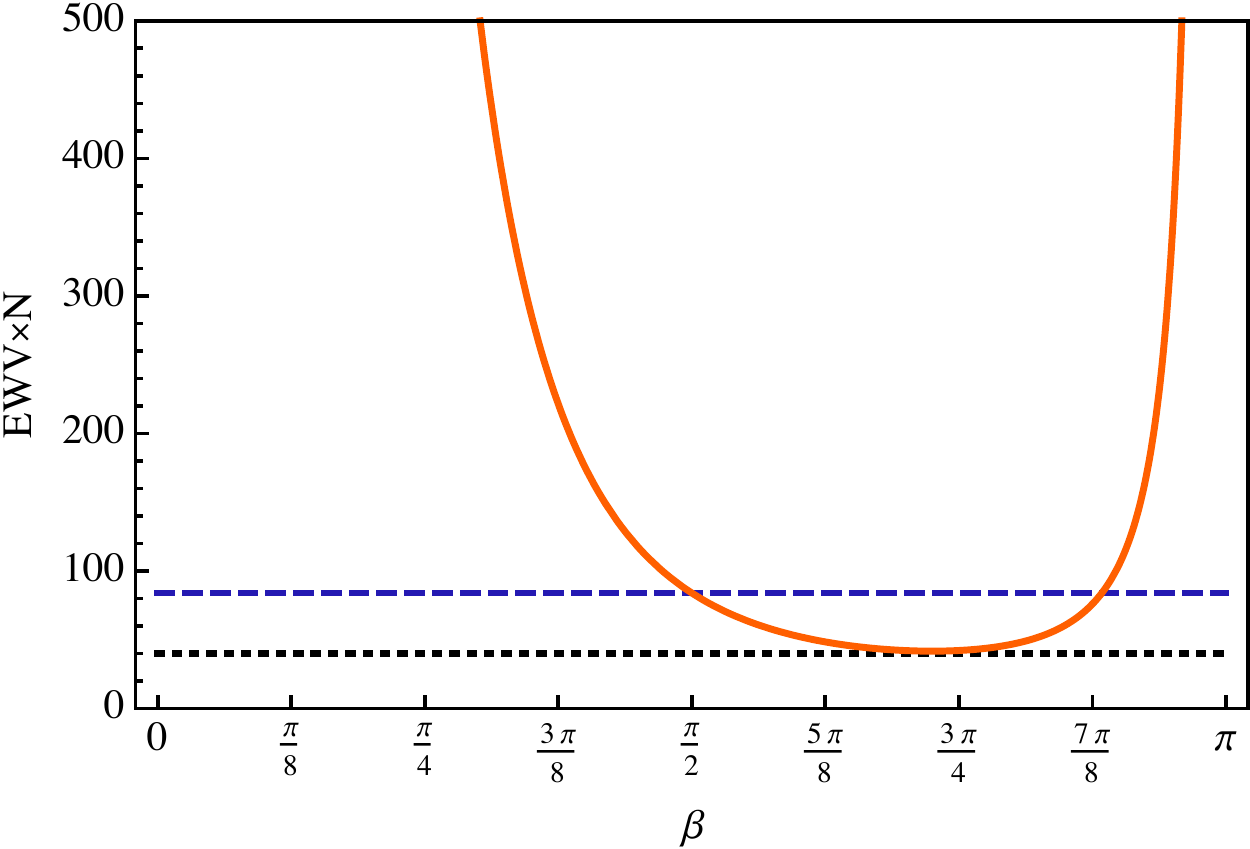}
\caption{ (Color online) The equally weighted variance (EWV) \cite{Sabatke2000} assesses the noise immunity of the wave-plate. Here we plot $\mathrm{EWV}\times N$ for one period as a function of $\beta$ (orange, solid), where $N$ is the total number of bins per period.  A smaller $\mathrm{EWV}$ is associated with better immunity to noise. The best noise immunity occurs at $\beta \approx 2.27 \approx 11\pi/15$ for  $\mathrm{EWV}\approx 41.7/N$. The black, dotted line shows $\mathrm{EWV}^{\mathrm{opt}}=40/N$, attainable by optimal tomographic schemes such as those that measure the Pauli matrices or the SIC-POVM \cite{Rehacek2004,Ling2006,Kalev2012a}. The blue, dashed line shows the EWV for a QWP ($\beta=\pi/2$), $\mathrm{EWV}^{\textsc{qwp}}=84/N$. Note that the EWV is dimensionless.}
\label{fig:Fig3}
\end{figure}

To characterize an $n$-qubit state, one measures a joint probability of detecting a photon in the $H$ mode of each PBS. This is given by
\begin{align}\label{eq:ProbabilityH}
p_{n}(t)={}&\frac{1}{2^{n}}\sum_{i_{1},  \dots, i_{n}=0}^{3} S_{i_{1} \dots,i_{n}}\chi_{1,i_{1}} \dots \chi_{n,i_{n}}\,,
\end{align}
where 
\begin{align}\label{eq:chi}
\chi_{{m},i} :=\mathrm{Tr}[ \hat \sigma _{{i}}\hat{M}_{m}^{H}(t)]\,,
\end{align} 
and therefore
\begin{subequations}
\begin{align}\label{eq:chiSpecific}
\chi _{m,0}={}&1\\
\chi _{m,1}={}&{s}^2 \sin \left(2 \omega_{m}{t}\right)\\
\chi  _{m,2}={}&2 {c} {s} \sin \left(\omega_{m}{t}\right)\\
\chi _{m,3}={}&{c}^2+{s}^2 \cos \left(2 \omega_{m}{t}\right)
\end{align}
\end{subequations}
where ${c}=\cos(\beta/2)$ and ${s}=\sin \left(\beta/2\right)$.

Note that the choice of analyzing the signal from mode $H$ rather than mode $V$ is arbitrary and typically both modes will need to be measured to ensure normalised probabilities.

Without loss of generality, we restrict $0<\omega_{1}<\dots<\omega_{n}$. For two qubits, $\omega_{2}={r}\omega_{1}$ where $r>1$. If $r$ is an irrational number, the signal does not have a finite period. If $r$ is a rational number, we can write ${r}={p}/{q}$, where $p$ and $q$ are integers. In this case, the period of the two-qubit signal is given by
\begin{align}
{T}({r})=\frac{2\pi q}{\omega_{1}\mathrm{gcd}(p,q)}\,,
\end{align}
where $\mathrm{gcd}(p,q)$ is the greatest common denominator of $p$ and $q$. For $n>2$, the period of the signal can be determined via recursion. A shorter period is favourable from an experimental perspective which, for a constant $\omega_{1}$, occurs when $r$ is an integer. The lowest integer that ensures sufficient Fourier coefficients to solve for Stokes' parameters is $r=5$.

In practice $p_{n}$  will not be a continuous function of time but rather a discretized approximation. The discretized signal will be divided into time bins, with $\mathcal{N}$ coincidence counts in each bin.  The number of time bins per period, $N$, must be at least the Nyquist rate, i.e. twice the highest frequency contained within the signal,   to avoid aliasing. 

The discrete probability in bin $\tau_{j}$ will be given by 
\begin{align}\label{eq:Probability1}
p_{n}(\tau_{j})=\frac{n_{H\dots H}}{\mathcal{N}}\,,
\end{align}
where $\mathcal{N}=\sum_{k_1,\dots,k_{2^{n}}=H,V} n_{k_1,\dots,k_{2^{n}}}$, $n$ is the number of qubits and $n_{k_1,\dots,k_{2^{n}}}$ is the number of coincidence counts for a given projector $\hat{M}_{m}^{k_1}(t)\otimes\dots\otimes\hat{M}_{m}^{k_{2^{n}}}(t)$. 

 In principle, $\beta$ can take on any value other than an integer multiple of $\pi$. However in practice, some values will be more susceptible to noise than others.  We use the equally weighted variance (EWV) \cite{Sabatke2000}, which assesses the noise immunity of the wave-plate, to show that $\beta\approx11\pi/15$ is most immune to noise, as defined in \cite{Sabatke2000}. A plot of the EWV is shown in FIG \ref{fig:Fig3}. Such a retardance can be achieved with an off-the-shelf wave plate designed for a wave length different to that of the experiment. 

In the remainder of this section, we provide specific examples for one- and two-qubit states. 

\begin{figure*}[t]
\includegraphics[width=0.85\textwidth]{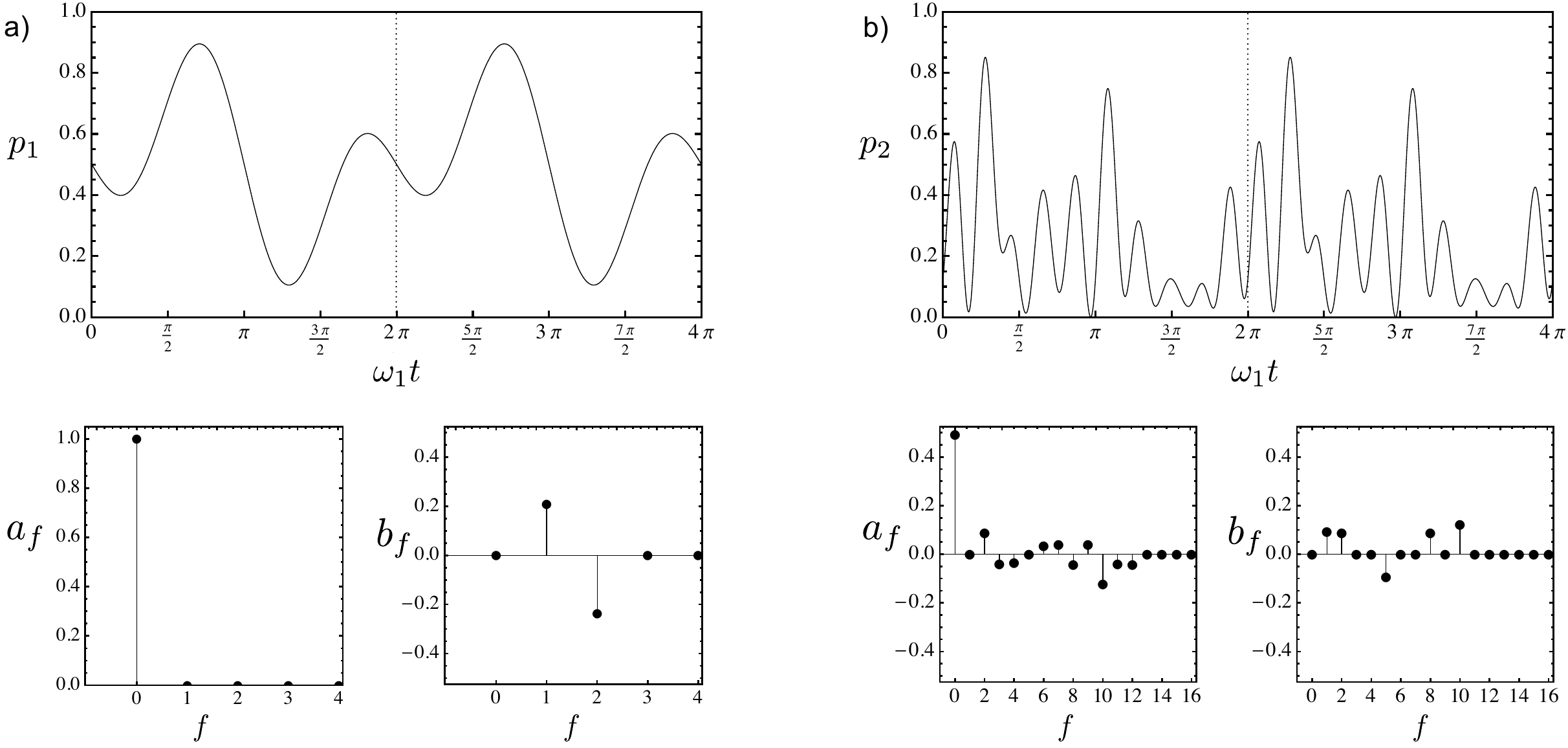}
\caption{The probability signal $p_{n}$, as a function of revolutions of the first wave plate, and Fourier coefficients ${a}_{f}$ and ${b}_{f}$ for the state: a) $\hat\rho_{1}$ given in Equation (\ref{eq:rho1example}); and b) $\ket{\psi}{2}$ given in Equation (\ref{eq:State2}). The label $f$ denotes the term in the Fourier series and corresponds to the subscript in Equations (\ref{eq:Probability1Fourier}) and (\ref{eq:113})}
\label{fig:Fig4}
\end{figure*}

\subsection{Example: one qubit}

For a single qubit, the signal is given by
\begin{align}
p_{1}(t)={}&\frac{1}{2}\sum_{{i}={0}}^{3} S_{{i}}\chi_{{1},{i}}\\\label{eq:Probability1Stokes}
\begin{split}
={}&\frac{{S}_0}{2} +\frac{ {S}_1}{2} {s}^2 \sin \left(2 \omega_{1}t\right)+{S}_2{c} {s}  \sin \left(\omega_{1}t\right)\\
&+\frac{{S}_3}{2}  \left({c}^2+{s}^2 \cos \left(2 \omega_{1}t\right)\right)\,.
   \end{split}
\end{align}
This can be written as
\begin{align}\label{eq:Probability1Fourier}
\begin{split}
p_{1}(t)\equiv{}&\frac{{a}_{0}}{2} +{b}_{1}\sin\left(\omega_{1}t\right)\\
&+{a}_{2} \cos\left(2\omega_{1}t\right)+{b}_{2} \sin\left(2\omega_{1}t\right)\,,
\end{split}
\end{align}
where the Fourier coefficients are given by
\begin{subequations}\label{eq:Fourier1Stokes}
\begin{align}
{a}_{0}={}&S_{0}+S_{3}{c}^2\,;\\
{b}_{1}={}&{S}_2{c} {s}\,;\\
{a}_{2}={}&\frac{{S}_3}{2} {s}^2\,;\\
{b}_{2}={}&\frac{ {S}_1}{2} {s}^2\,,
\end{align}
\end{subequations}
where ${c}=\cos(\beta/2)$ and ${s}=\sin \left(\beta/2\right)$. Linear inversion of Equations (\ref{eq:Fourier1Stokes}) gives the Stokes' parameters in terms of the Fourier coefficients:
\begin{subequations}\label{eq:Stokes1Fourier}
\begin{align}
{S}_{0}={}& a_0- \frac{2 a_2{c}^2}{{s}^2}\,;\\
{S}_{1}={}& \frac{2 b_2}{{s}^2}\,;\\
{S}_{2}={}& \frac{b_1}{{c}   {s}}\,;\\
{S}_{3}={}&\frac{2 a_2}{{s}^2} \,.
\end{align}
\end{subequations}
Substitution into Equation (\ref{eq:RhoGeneral}), gives the density matrix in terms of the Fourier coefficients:
\begin{align}\label{eq:dens}
\rho_{1}={}&\left(
\begin{array}{cc}
 \frac{1}{2} \left(a_0+2 a_2\right) & \frac{b_2}{{s}^2}-i\frac{ b_1}{2 {c} {s}} \\
\frac{b_2}{{s}^2}+ i\frac{ b_1}{2 {c} {s}} & \frac{1}{2} \left(a_0+2 a_2\right)-\frac{2 a_2}{{s}^2}
\end{array}
\right)\,.
\end{align}

As an example, consider a single-qubit state $\ket{\psi}{1}$ that has experienced depolarizing noise, characterized by the parameter $d$, such that
\begin{align}\label{eq:rho1example}
\hat\rho_{1}=d\ket{\psi}{1}\bra{\psi}{1}+(1-d)\hat\sigma_{z}\ket{\psi}{1}\bra{\psi}{1}\hat\sigma_{z}\,.
\end{align}
Specifically, let's consider
\begin{align}
\ket{\psi}{1}=\frac{1}{\sqrt{2}}(\ket{H}{}+\ee{-i\pi/4}\ket{V}{})
\end{align} 
and $d=0.1$. A retardance of $\beta=11\pi/15$ produces the signal shown in FIG \ref{fig:Fig4} a). Performing a fast Fourier transform (FFT) of the discretized signal yields the Fourier coefficients in FIG \ref{fig:Fig4} a). The coefficients ${a}_{f}$ and ${b}_{f}$ correspond to the real and imaginary parts of the list generated by the FFT respectively. Inserting the coefficients, ${a}_{0}=1$, ${b}_{1}=0.210$, ${a}_{2}=0$ and ${b}_{2}=-0.236$ into the density matrix in Equation (\ref{eq:dens}) gives
\begin{align}
\rho_{1}={}&\left(
\begin{array}{cc}
 0.5 & -0.283 -0.283 i \\
 -0.283 +0.283 i & 0.5
\end{array}
\right)\,,
\end{align}
which corresponds to the density operator in Equation (\ref{eq:rho1example}) for $d=0.1$.

\subsection{Example: two qubits}\label{sec:TwoQubits}

For two qubits, the joint probability of detecting a photon in the horizontal output ports of each PBS is given by
\begin{align}\label{eq:Probability2}
p_{2}(t)={}&\sum_{{i}_{1},{i}_{2}={0}}^{3} \frac{S_{{i}_{1},{i}_{2}}}{4}\chi_{{1},{i}_{1}}\chi_{{2},{i}_{2}}\\\label{eq:113}
={}&\frac{a_{0}}{2}+\sum_{f=1}\left({a}_{f}\cos(\omega'_{f}t)+{b}_{f}\sin(\omega'_{f}t)\right)\,,
\end{align}
where in the second line, we have written the signal in terms of its Fourier coefficients. The extent of the summation depends on the specific choice of relative frequencies, and $\omega'_{f}$ are functions of $\omega_1$ and $\omega_2$ from the set $\{\omega_{1},\omega_{2},2 \omega_{1},2 \omega_{2},\omega_{1}\pm\omega_{2},\omega_{1}\pm2 \omega_{2},2 \omega_{1}\pm\omega  _2,2 \omega_{1}\pm2 \omega   _2\}$. The elements of this set are not necessarily in order of size, and to relate them to $\omega'_{f}$ one needs to consider explicit values for $\omega_{1}$ and $\omega_{2}$.

As an example, consider the two-qubit state
\begin{align}\label{eq:State2}
\ket{\psi}{2}=\frac{1}{\sqrt{2}}\left(\ket{H}{}\ket{V}{}+\ket{R}{}\ket{L}{}\right)\,.
\end{align}

Wave plate retardances of $\beta=11\pi/15$ and a frequency ratio ${r}=\omega_{2}/\omega_{1}=5$ produces the signal shown in FIG \ref{fig:Fig4}. The explicit expression for ${p}_{2}(t)$ for this choice of measurement settings can be found in Appendix \ref{sec:ProbabilitySignal}. Performing a fast Fourier transform (FFT) of the discretized signal yields the Fourier coefficients in FIG \ref{fig:Fig4} b). 

Linear inversion of the expressions for the Fourier coefficients in terms of the Stokes' parameters, given in Equations (\ref{eq:Fourier2a}), followed by substitution into Equation (\ref{eq:RhoGeneral}), along with the Fourier coefficients determined from the signal in FIG \ref{fig:Fig4} b), gives the reconstructed density matrix
\begin{widetext}
\begin{align}\label{eq:rho2example}
\rho_{2}={}&\left(
\begin{array}{cccc}
 0.125 & 0.25 +0.125 i & -0.125 i & 0.125 \\
 0.25 -0.125 i & 0.625 & -0.125-0.25 i & 0.25 -0.125 i \\
 0.125 i & -0.125+0.25 i & 0.125 & 0.125 i \\
 0.125 & 0.25 +0.125 i & -0.125 i & 0.125
\end{array}
\right)\,,
\end{align}
\end{widetext}
which corresponds to the density operator $\hat\rho_{2}=\ket{\psi}{2}\bra{\psi}{2}$ where $\ket{\psi}{2}$ is defined in Equation (\ref{eq:State2}). In general, given separable qubits, ${p}_{2}$ factorizes into a product of ${p}_{1}$ for each qubit. 

\section{Summary \& concluding remarks}\label{sec:Conclusion}

We presented a scheme for performing quantum state tomography of photonic  polarization-encoded multi-qubit states. The scheme is simpler than standard tomographic protocols in that only one wave plate and one polarizing beam splitter is required per photon mode. 

In this scheme, photon-counting detectors measure a pseudo-continuous time-dependent joint probability as the wave plates rotate at frequency $\Omega_{m}$. The Fourier coefficients of the signal give the Stokes' parameters which describe the state. For a single qubit, the optimal wave plate retardance is $\beta\approx 11\pi/15$. 

This technique reduces the number of required optical elements and the experimental complexity scales linearly with the number of qubits, in terms of the number of settings required (wave plate rotation frequencies) rather than exponentially, as is the case with QST that uses discrete measurement settings.

An open question is whether the representation of a quantum state as a continuous signal will provide intuitive means for establishing certain properties of the state such as its entanglement. 

\section{Acknowledgements}
The authors thank Dylan Mahler and Paul Kwiat for helpful discussions. This work was funded by NSERC (USRA) and the DARPA (QuBE) program. 

\appendix

\section{Probability signal and Fourier coefficients for two-qubit state}\label{sec:ProbabilitySignal}

In this appendix, we give the probability signal for the specific two-qubit example described in Section \ref{sec:TwoQubits}. We also provide expressions for the Fourier coefficients in terms of the Stokes' parameters, as well as the inverted expressions for the Stokes' parameters in terms of the Fourier coefficients. 

The signal probability for the state
\begin{align}\label{eq:state2exampleB}
\ket{\psi}{2}=\frac{1}{\sqrt{2}}\left(\ket{H}{}\ket{V}{}+\ket{R}{}\ket{L}{}\right)\,,
\end{align}
with a frequency ratio ${r}=\omega_{2}/\omega_{1}=5$,  is 
\begin{align}\label{eq:Probability2}
\begin{split}
p_{2}({t}) ={}&\frac{{a}_{0}}{2}+{b}_{1} \sin \left(\omega_{1}{t}\right)+{b}_{2} \sin \left(2\omega_{1}{t}\right)+{b}_{3}\sin \left(3\omega_{1}{t}\right)\\
&+{b}_{5} \sin \left(5\omega_{1}{t}\right)+   {b}_{7}\sin \left(7\omega_{1}{t}\right)+{b}_{8}\sin \left(8\omega_{1}{t}\right)\\
&+  {b}_{9}\sin \left(9\omega_{1}{t}\right)+{b}_{10} \sin \left(10\omega_{1}{t}\right)+{b}_{11}\sin \left(11\omega_{1}{t}\right)\\
&+{b}_{12}\sin \left(12\omega_{1}{t}\right)  +{a}_{2} \cos \left(2\omega_{1}{t}\right) +{a}_{3}  \cos \left(3\omega_{1}{t}\right)\\
  &+{a}_{8}\cos \left(8\omega_{1}{t}\right)+ {a}_{4}\cos \left(4\omega_{1}{t}\right)+ {a}_{6}\cos \left(6\omega_{1}{t}\right)\\
&+ {a}_{7} \cos  \left(7\omega_{1}{t}\right)+  {a}_{9}\cos \left(9\omega_{1}{t}\right)+{a}_{10} \cos \left(10\omega_{1}{t}\right)\\
&+ {a}_{11}\cos \left(11\omega_{1}{t}\right)+ {a}_{12}\cos \left(12\omega_{1}{t}\right)\,,
\end{split}
 \end{align}
 where the Fourier coefficients are
\begin{subequations}\label{eq:Fourier2a}
\begin{align}
 a_0={}&  \left(c^2 \left(c^2 {S}_{3,3}+{S}_{0,3}+{S}_{3,0}\right)+{S}_{0,0}\right)/2 \\
 a_2={}&s^2 \left(c^2 {S}_{3,3}+{S}_{3,0}\right)/4 \\
 a_3={}&-a_7= c s^3 {S}_{1,2}/4 \\
 a_4={}& -a_6=c^2 s^2 {S}_{2,2}/2 \\
 a_8={}& s^4 \left({S}_{1,1}+{S}_{3,3}\right) /8\\
 a_9={}&=-a_{11}= c s^3 {S}_{2,1} /4\\
 a_{10}={}& s^2 \left(c^2 {S}_{3,3}+{S}_{0,3}\right) /4\\
 a_{12}={}&s^4 \left({S}_{3,3}-{S}_{1,1}\right)/8\,,
  \end{align}
\end{subequations}
and
\begin{subequations}\label{eq:Fourier2b}
\begin{align}
b_1={}&  c s \left(c^2 {S}_{2,3}+{S}_{2,0}\right) /2\\
 b_2={}&  s^2 \left(c^2 {S}_{1,3}+{S}_{1,0}\right) /2\\
 b_3={}& b_7= c s^3 {S}_{3,2} /4\\
 b_5={}&  c s \left(c^2 {S}_{3,2}+{S}_{0,2}\right)/2 \\
 b_8={}&  s^4 \left({S}_{3,1}-{S}_{1,3}\right) /8\\
 b_9={}& -b_{11}=- c s^3 {S}_{2,3}/4 \\
 b_{10}={}&  s^2 \left(c^2 {S}_{3,1}+{S}_{0,1}\right) /4\\
 b_{12}={}&  s^4 \left({S}_{1,3}+{S}_{3,1}\right)/8\,,
\end{align}
\end{subequations}
where ${c}=\cos(\beta/2)$ and ${s}=\sin \left(\beta/2\right)$. Inverting Equations (\ref{eq:Fourier2a}) and (\ref{eq:Fourier2b}) gives
\begin{subequations}\label{eq:Stokes2Fourier}
\begin{align}
 {S}_{0,0}={}&{4 c^2 \left(\left(a_8+a_{12}\right) c^2-\left(a_2+a_{10}\right) s^2\right)}/{s^4}+2 a_0 \\
 {S}_{0,1}={}& -{4 \left(\left(b_8+b_{12}\right) c^2-b_{10} s^2\right)}/{s^4} \\
 {S}_{0,2}={}&{2 b_5}/{c s}-{4 b_3 c}/{s^3} \\
 {S}_{0,3}={}&{4 a_{10}}/{s^2}-{4 \left(a_8+a_{12}\right) c^2}/{s^4} \\
 {S}_{1,0}={}&{4 \left(\left(b_8-b_{12}\right) c^2+b_2 s^2\right)}/{s^4} \\
 {S}_{1,1}={}&{4 \left(a_8-a_{12}\right)}/{s^4} \\
 {S}_{1,2}={}&{4 a_3}/{c s^3} \\
 {S}_{1,3}={}& -{4 \left(b_8-b_{12}\right)}/{s^4} \\
 {S}_{2,0}={}&{4 b_9 c}/{s^3}+{2 b_1}/{c s} \\
 {S}_{2,1}={}&{4 a_9}/{c s^3} \\
 {S}_{2,2}={}&{2 a_4}/{c^2 s^2} \\
 {S}_{2,3}={}& -{4 b_9}/{c s^3} \\
 {S}_{3,0}={}&{4 a_2}/{s^2}-{4 \left(a_8+a_{12}\right) c^2}/{s^4} \\
 {S}_{3,1}={}&{4 \left(b_8+b_{12}\right)}/{s^4} \\
 {S}_{3,2}={}&{4 b_3}/{c s^3} \\
 {S}_{3,3}={}&{4 \left(a_8+a_{12}\right)}/{s^4}\,.
\end{align}
\end{subequations}
Substituting the Stokes' parameters, along with the Fourier coefficients in FIG \ref{fig:Fig4} b), into Equation (\ref{eq:RhoGeneral}), gives the reconstructed density matrix in Equation (\ref{eq:rho2example}) which corresponds to the density operator $\hat\rho_{2}=\ket{\psi}{2}\bra{\psi}{2}$ where $\ket{\psi}{2}$ is defined in Equation (\ref{eq:state2exampleB}).


\begin{thebibliography}{42}%
\makeatletter
\providecommand \@ifxundefined [1]{%
 \@ifx{#1\undefined}
}%
\providecommand \@ifnum [1]{%
 \ifnum #1\expandafter \@firstoftwo
 \else \expandafter \@secondoftwo
 \fi
}%
\providecommand \@ifx [1]{%
 \ifx #1\expandafter \@firstoftwo
 \else \expandafter \@secondoftwo
 \fi
}%
\providecommand \natexlab [1]{#1}%
\providecommand \enquote  [1]{``#1''}%
\providecommand \bibnamefont  [1]{#1}%
\providecommand \bibfnamefont [1]{#1}%
\providecommand \citenamefont [1]{#1}%
\providecommand \href@noop [0]{\@secondoftwo}%
\providecommand \href [0]{\begingroup \@sanitize@url \@href}%
\providecommand \@href[1]{\@@startlink{#1}\@@href}%
\providecommand \@@href[1]{\endgroup#1\@@endlink}%
\providecommand \@sanitize@url [0]{\catcode `\\12\catcode `\$12\catcode
  `\&12\catcode `\#12\catcode `\^12\catcode `\_12\catcode `\%12\relax}%
\providecommand \@@startlink[1]{}%
\providecommand \@@endlink[0]{}%
\providecommand \url  [0]{\begingroup\@sanitize@url \@url }%
\providecommand \@url [1]{\endgroup\@href {#1}{\urlprefix }}%
\providecommand \urlprefix  [0]{URL }%
\providecommand \Eprint [0]{\href }%
\@ifxundefined \urlstyle {%
  \providecommand \doi  [0]{\begingroup \@sanitize@url \@doi}%
  \providecommand \@doi [1]{\endgroup \@@startlink {\doibase
  #1}doi:\discretionary {}{}{}#1\@@endlink }%
}{%
  \providecommand \doi  [0]{doi:\discretionary{}{}{}\begingroup
  \urlstyle{rm}\Url }%
}%
\providecommand \doibase [0]{http://dx.doi.org/}%
\providecommand \Doi [0]{\begingroup \@sanitize@url \@Doi }%
\providecommand \@Doi  [1]{\endgroup\@@startlink{\doibase#1}\@@Doi}%
\providecommand \@@Doi [1]{#1\@@endlink}%
\providecommand \selectlanguage [0]{\@gobble}%
\providecommand \bibinfo  [0]{\@secondoftwo}%
\providecommand \bibfield  [0]{\@secondoftwo}%
\providecommand \translation [1]{[#1]}%
\providecommand \BibitemOpen [0]{}%
\providecommand \bibitemStop [0]{}%
\providecommand \bibitemNoStop [0]{.\EOS\space}%
\providecommand \EOS [0]{\spacefactor3000\relax}%
\providecommand \BibitemShut  [1]{\csname bibitem#1\endcsname}%
\bibitem [{\citenamefont {Kok}\ and\ \citenamefont {Lovett}(2010)}]{Kok2010}%
  \BibitemOpen
  \bibfield  {author} {\bibinfo {author} {\bibfnamefont {P.}~\bibnamefont
  {Kok}}\ and\ \bibinfo {author} {\bibfnamefont {B.~W.}\ \bibnamefont
  {Lovett}},\ }\href@noop {} {\emph {\bibinfo {title} {Introduction to Optical
  Quantum Information Processing}}},\ \bibinfo {edition} {1st}\ ed.\ (\bibinfo
  {publisher} {Cambridge University Press},\ \bibinfo {year}
  {2010})\BibitemShut {NoStop}%
\bibitem [{\citenamefont {Bennett}\ and\ \citenamefont
  {Brassard}(1984)}]{Bennett1984}%
  \BibitemOpen
  \bibfield  {author} {\bibinfo {author} {\bibfnamefont {C.~H.}\ \bibnamefont
  {Bennett}}\ and\ \bibinfo {author} {\bibfnamefont {G.}~\bibnamefont
  {Brassard}},\ }in\ \href@noop {} {\emph {\bibinfo {booktitle} {Proceedings of
  {IEEE} {I}nternational {C}onference on {C}omputers, {S}ystems and {S}ignal
  {P}rocessing}}}\ (\bibinfo  {publisher} {IEEE},\ \bibinfo {address} {New
  York},\ \bibinfo {year} {1984})\ pp.\ \bibinfo {pages} {175--179},\ \bibinfo
  {note} {{B}angalore, India, December 1984}\BibitemShut {NoStop}%
\bibitem [{\citenamefont {Nielsen}\ and\ \citenamefont
  {Chuang}(2000)}]{Nielsen2000}%
  \BibitemOpen
  \bibfield  {author} {\bibinfo {author} {\bibfnamefont {M.~A.}\ \bibnamefont
  {Nielsen}}\ and\ \bibinfo {author} {\bibfnamefont {I.~L.}\ \bibnamefont
  {Chuang}},\ }\href@noop {} {\emph {\bibinfo {title} {Quantum computation and
  quantum information}}}\ (\bibinfo  {publisher} {Cambridge University Press},\
  \bibinfo {address} {Cambridge},\ \bibinfo {year} {2000})\BibitemShut
  {NoStop}%
\bibitem [{\citenamefont {Vogel}\ and\ \citenamefont
  {Risken}(1989)}]{Vogel1989}%
  \BibitemOpen
  \bibfield  {author} {\bibinfo {author} {\bibfnamefont {K.}~\bibnamefont
  {Vogel}}\ and\ \bibinfo {author} {\bibfnamefont {H.}~\bibnamefont {Risken}},\
  }\Doi {10.1103/PhysRevA.40.2847} {\bibfield  {journal} {\bibinfo  {journal}
  {Phys. Rev. A},\ }\textbf {\bibinfo {volume} {40}},\ \bibinfo {pages} {2847}
  (\bibinfo {year} {1989})}\BibitemShut {NoStop}%
\bibitem [{\citenamefont {Leonhardt}(1995)}]{Leonhardt1995}%
  \BibitemOpen
  \bibfield  {author} {\bibinfo {author} {\bibfnamefont {U.}~\bibnamefont
  {Leonhardt}},\ }\Doi {10.1103/PhysRevLett.74.4101} {\bibfield  {journal}
  {\bibinfo  {journal} {Phys. Rev. Lett.},\ }\textbf {\bibinfo {volume} {74}},\
  \bibinfo {pages} {4101} (\bibinfo {year} {1995})}\BibitemShut {NoStop}%
\bibitem [{\citenamefont {James}\ \emph {et~al.}(2001)\citenamefont {James},
  \citenamefont {Kwiat}, \citenamefont {Munro},\ and\ \citenamefont
  {White}}]{James2001}%
  \BibitemOpen
  \bibfield  {author} {\bibinfo {author} {\bibfnamefont {D.~F.~V.}\
  \bibnamefont {James}}, \bibinfo {author} {\bibfnamefont {P.~G.}\ \bibnamefont
  {Kwiat}}, \bibinfo {author} {\bibfnamefont {W.~J.}\ \bibnamefont {Munro}}, \
  and\ \bibinfo {author} {\bibfnamefont {A.~G.}\ \bibnamefont {White}},\ }\Doi
  {10.1103/PhysRevA.64.052312} {\bibfield  {journal} {\bibinfo  {journal}
  {Phys. Rev. A},\ }\textbf {\bibinfo {volume} {64}},\ \bibinfo {pages}
  {052312} (\bibinfo {year} {2001})}\BibitemShut {NoStop}%
\bibitem [{\citenamefont {White}\ \emph {et~al.}(1999)\citenamefont {White},
  \citenamefont {James}, \citenamefont {Eberhard},\ and\ \citenamefont
  {Kwiat}}]{White1999}%
  \BibitemOpen
  \bibfield  {author} {\bibinfo {author} {\bibfnamefont {A.~G.}\ \bibnamefont
  {White}}, \bibinfo {author} {\bibfnamefont {D.~F.~V.}\ \bibnamefont {James}},
  \bibinfo {author} {\bibfnamefont {P.~H.}\ \bibnamefont {Eberhard}}, \ and\
  \bibinfo {author} {\bibfnamefont {P.~G.}\ \bibnamefont {Kwiat}},\ }\Doi
  {10.1103/PhysRevLett.83.3103} {\bibfield  {journal} {\bibinfo  {journal}
  {Phys. Rev. Lett.},\ }\textbf {\bibinfo {volume} {83}},\ \bibinfo {pages}
  {3103} (\bibinfo {year} {1999})}\BibitemShut {NoStop}%
\bibitem [{\citenamefont {Stokes}(1852)}]{Stokes1852}%
  \BibitemOpen
  \bibfield  {author} {\bibinfo {author} {\bibfnamefont {G.~C.}\ \bibnamefont
  {Stokes}},\ }\href@noop {} {\bibfield  {journal} {\bibinfo  {journal}
  {Cambridge Philos. Soc},\ }\textbf {\bibinfo {volume} {9}},\ \bibinfo {pages}
  {399} (\bibinfo {year} {1852})}\BibitemShut {NoStop}%
\bibitem [{\citenamefont {de~Burgh}\ \emph {et~al.}(2008)\citenamefont
  {de~Burgh}, \citenamefont {Langford}, \citenamefont {Doherty},\ and\
  \citenamefont {Gilchrist}}]{deBurgh2008}%
  \BibitemOpen
  \bibfield  {author} {\bibinfo {author} {\bibfnamefont {M.~D.}\ \bibnamefont
  {de~Burgh}}, \bibinfo {author} {\bibfnamefont {N.~K.}\ \bibnamefont
  {Langford}}, \bibinfo {author} {\bibfnamefont {A.~C.}\ \bibnamefont
  {Doherty}}, \ and\ \bibinfo {author} {\bibfnamefont {A.}~\bibnamefont
  {Gilchrist}},\ }\Doi {10.1103/PhysRevA.78.052122} {\bibfield  {journal}
  {\bibinfo  {journal} {Phys. Rev. A},\ }\textbf {\bibinfo {volume} {78}},\
  \bibinfo {pages} {052122} (\bibinfo {year} {2008})}\BibitemShut {NoStop}%
\bibitem [{\citenamefont {{\v R}eh{\'a}{\v c}ek}\ \emph
  {et~al.}(2004)\citenamefont {{\v R}eh{\'a}{\v c}ek}, \citenamefont
  {Englert},\ and\ \citenamefont {Kaszlikowski}}]{Rehacek2004}%
  \BibitemOpen
  \bibfield  {author} {\bibinfo {author} {\bibfnamefont {J.}~\bibnamefont {{\v
  R}eh{\'a}{\v c}ek}}, \bibinfo {author} {\bibfnamefont {B.-G.}\ \bibnamefont
  {Englert}}, \ and\ \bibinfo {author} {\bibfnamefont {D.}~\bibnamefont
  {Kaszlikowski}},\ }\Doi {10.1103/PhysRevA.70.052321} {\bibfield  {journal}
  {\bibinfo  {journal} {Phys. Rev. A},\ }\textbf {\bibinfo {volume} {70}},\
  \bibinfo {pages} {052321} (\bibinfo {year} {2004})}\BibitemShut {NoStop}%
\bibitem [{\citenamefont {Ling}\ \emph {et~al.}(2006)\citenamefont {Ling},
  \citenamefont {Soh}, \citenamefont {Lamas-Linares},\ and\ \citenamefont
  {Kurtsiefer}}]{Ling2006}%
  \BibitemOpen
  \bibfield  {author} {\bibinfo {author} {\bibfnamefont {A.}~\bibnamefont
  {Ling}}, \bibinfo {author} {\bibfnamefont {K.~P.}\ \bibnamefont {Soh}},
  \bibinfo {author} {\bibfnamefont {A.}~\bibnamefont {Lamas-Linares}}, \ and\
  \bibinfo {author} {\bibfnamefont {C.}~\bibnamefont {Kurtsiefer}},\ }\Doi
  {10.1103/PhysRevA.74.022309} {\bibfield  {journal} {\bibinfo  {journal}
  {Phys. Rev. A},\ }\textbf {\bibinfo {volume} {74}},\ \bibinfo {pages}
  {022309} (\bibinfo {year} {2006})}\BibitemShut {NoStop}%
\bibitem [{\citenamefont {Medendorp}\ \emph {et~al.}(2011)\citenamefont
  {Medendorp}, \citenamefont {Torres-Ruiz}, \citenamefont {Shalm},
  \citenamefont {Tabia}, \citenamefont {Fuchs},\ and\ \citenamefont
  {Steinberg}}]{Medendorp2011}%
  \BibitemOpen
  \bibfield  {author} {\bibinfo {author} {\bibfnamefont {Z.~E.~D.}\
  \bibnamefont {Medendorp}}, \bibinfo {author} {\bibfnamefont {F.~A.}\
  \bibnamefont {Torres-Ruiz}}, \bibinfo {author} {\bibfnamefont {L.~K.}\
  \bibnamefont {Shalm}}, \bibinfo {author} {\bibfnamefont {G.~N.~M.}\
  \bibnamefont {Tabia}}, \bibinfo {author} {\bibfnamefont {C.~A.}\ \bibnamefont
  {Fuchs}}, \ and\ \bibinfo {author} {\bibfnamefont {A.~M.}\ \bibnamefont
  {Steinberg}},\ }\Doi {10.1103/PhysRevA.83.051801} {\bibfield  {journal}
  {\bibinfo  {journal} {Phys. Rev. A},\ }\textbf {\bibinfo {volume} {83}},\
  \bibinfo {pages} {051801} (\bibinfo {year} {2011})}\BibitemShut {NoStop}%
\bibitem [{\citenamefont {Kalev}\ \emph
  {et~al.}(2012){\natexlab{a}}\citenamefont {Kalev}, \citenamefont {Shang},\
  and\ \citenamefont {Englert}}]{Kalev2012a}%
  \BibitemOpen
  \bibfield  {author} {\bibinfo {author} {\bibfnamefont {A.}~\bibnamefont
  {Kalev}}, \bibinfo {author} {\bibfnamefont {J.}~\bibnamefont {Shang}}, \ and\
  \bibinfo {author} {\bibfnamefont {B.-G.}\ \bibnamefont {Englert}},\ }\Doi
  {10.1103/PhysRevA.85.052116} {\bibfield  {journal} {\bibinfo  {journal}
  {Phys. Rev. A},\ }\textbf {\bibinfo {volume} {85}},\ \bibinfo {pages}
  {052116} (\bibinfo {year} {2012}{\natexlab{a}})}\BibitemShut {NoStop}%
\bibitem [{\citenamefont {Kosut}\ \emph {et~al.}(2004)\citenamefont {Kosut},
  \citenamefont {Walmsley},\ and\ \citenamefont {Rabitz}}]{Kosut2004}%
  \BibitemOpen
  \bibfield  {author} {\bibinfo {author} {\bibfnamefont {R.}~\bibnamefont
  {Kosut}}, \bibinfo {author} {\bibfnamefont {I.~A.}\ \bibnamefont {Walmsley}},
  \ and\ \bibinfo {author} {\bibfnamefont {H.}~\bibnamefont {Rabitz}},\
  }\href@noop {} {\bibfield  {journal} {\bibinfo  {journal}
  {arXiv:quant-ph/0411093v1}} (\bibinfo {year} {2004})}\BibitemShut {NoStop}%
\bibitem [{\citenamefont {Nunn}\ \emph {et~al.}(2010)\citenamefont {Nunn},
  \citenamefont {Smith}, \citenamefont {Puentes}, \citenamefont {Walmsley},\
  and\ \citenamefont {Lundeen}}]{Nunn2010}%
  \BibitemOpen
  \bibfield  {author} {\bibinfo {author} {\bibfnamefont {J.}~\bibnamefont
  {Nunn}}, \bibinfo {author} {\bibfnamefont {B.~J.}\ \bibnamefont {Smith}},
  \bibinfo {author} {\bibfnamefont {G.}~\bibnamefont {Puentes}}, \bibinfo
  {author} {\bibfnamefont {I.~A.}\ \bibnamefont {Walmsley}}, \ and\ \bibinfo
  {author} {\bibfnamefont {J.~S.}\ \bibnamefont {Lundeen}},\ }\Doi
  {10.1103/PhysRevA.81.042109} {\bibfield  {journal} {\bibinfo  {journal}
  {Phys. Rev. A},\ }\textbf {\bibinfo {volume} {81}},\ \bibinfo {pages}
  {042109} (\bibinfo {year} {2010})}\BibitemShut {NoStop}%
\bibitem [{\citenamefont {Adamson}\ \emph {et~al.}(2007)\citenamefont
  {Adamson}, \citenamefont {Shalm}, \citenamefont {Mitchell},\ and\
  \citenamefont {Steinberg}}]{Adamson2007}%
  \BibitemOpen
  \bibfield  {author} {\bibinfo {author} {\bibfnamefont {R.~B.~A.}\
  \bibnamefont {Adamson}}, \bibinfo {author} {\bibfnamefont {L.~K.}\
  \bibnamefont {Shalm}}, \bibinfo {author} {\bibfnamefont {M.~W.}\ \bibnamefont
  {Mitchell}}, \ and\ \bibinfo {author} {\bibfnamefont {A.~M.}\ \bibnamefont
  {Steinberg}},\ }\Doi {10.1103/PhysRevLett.98.043601} {\bibfield  {journal}
  {\bibinfo  {journal} {Phys. Rev. Lett.},\ }\textbf {\bibinfo {volume} {98}},\
  \bibinfo {pages} {043601} (\bibinfo {year} {2007})}\BibitemShut {NoStop}%
\bibitem [{\citenamefont {Adamson}\ \emph {et~al.}(2008)\citenamefont
  {Adamson}, \citenamefont {Turner}, \citenamefont {Mitchell},\ and\
  \citenamefont {Steinberg}}]{Adamson2008}%
  \BibitemOpen
  \bibfield  {author} {\bibinfo {author} {\bibfnamefont {R.~B.~A.}\
  \bibnamefont {Adamson}}, \bibinfo {author} {\bibfnamefont {P.~S.}\
  \bibnamefont {Turner}}, \bibinfo {author} {\bibfnamefont {M.~W.}\
  \bibnamefont {Mitchell}}, \ and\ \bibinfo {author} {\bibfnamefont {A.~M.}\
  \bibnamefont {Steinberg}},\ }\Doi {10.1103/PhysRevA.78.033832} {\bibfield
  {journal} {\bibinfo  {journal} {Phys. Rev. A},\ }\textbf {\bibinfo {volume}
  {78}},\ \bibinfo {pages} {033832} (\bibinfo {year} {2008})}\BibitemShut
  {NoStop}%
\bibitem [{\citenamefont {Asorey}\ \emph {et~al.}(2011)\citenamefont {Asorey},
  \citenamefont {Facchi}, \citenamefont {Florio}, \citenamefont {Man'ko},
  \citenamefont {Marmo}, \citenamefont {Pascazio},\ and\ \citenamefont
  {Sudarshan}}]{Asorey2011}%
  \BibitemOpen
  \bibfield  {author} {\bibinfo {author} {\bibfnamefont {M.}~\bibnamefont
  {Asorey}}, \bibinfo {author} {\bibfnamefont {P.}~\bibnamefont {Facchi}},
  \bibinfo {author} {\bibfnamefont {G.}~\bibnamefont {Florio}}, \bibinfo
  {author} {\bibfnamefont {V.}~\bibnamefont {Man'ko}}, \bibinfo {author}
  {\bibfnamefont {G.}~\bibnamefont {Marmo}}, \bibinfo {author} {\bibfnamefont
  {S.}~\bibnamefont {Pascazio}}, \ and\ \bibinfo {author} {\bibfnamefont
  {E.}~\bibnamefont {Sudarshan}},\ }\Doi {10.1016/j.physleta.2010.12.056}
  {\bibfield  {journal} {\bibinfo  {journal} {Physics Letters A},\ }\textbf
  {\bibinfo {volume} {375}},\ \bibinfo {pages} {861 } (\bibinfo {year}
  {2011})},\ ISSN \bibinfo {issn} {0375-9601}\BibitemShut {NoStop}%
\bibitem [{\citenamefont {Bra\'{n}czyk}\ \emph {et~al.}(2012)\citenamefont
  {Bra\'{n}czyk}, \citenamefont {Mahler}, \citenamefont {Rozema}, \citenamefont
  {Darabi}, \citenamefont {Steinberg},\ and\ \citenamefont
  {James}}]{Branczyk2012a}%
  \BibitemOpen
  \bibfield  {author} {\bibinfo {author} {\bibfnamefont {A.~M.}\ \bibnamefont
  {Bra\'{n}czyk}}, \bibinfo {author} {\bibfnamefont {D.~H.}\ \bibnamefont
  {Mahler}}, \bibinfo {author} {\bibfnamefont {L.~A.}\ \bibnamefont {Rozema}},
  \bibinfo {author} {\bibfnamefont {A.}~\bibnamefont {Darabi}}, \bibinfo
  {author} {\bibfnamefont {A.~M.}\ \bibnamefont {Steinberg}}, \ and\ \bibinfo
  {author} {\bibfnamefont {D.~F.~V.}\ \bibnamefont {James}},\ }\href
  {http://stacks.iop.org/1367-2630/14/i=8/a=085003} {\bibfield  {journal}
  {\bibinfo  {journal} {New Journal of Physics},\ }\textbf {\bibinfo {volume}
  {14}},\ \bibinfo {pages} {085003} (\bibinfo {year} {2012})}\BibitemShut
  {NoStop}%
\bibitem [{\citenamefont {Mogilevtsev}\ \emph {et~al.}(2012)\citenamefont
  {Mogilevtsev}, \citenamefont {{\v R}eh{\'a}{\v c}ek},\ and\ \citenamefont
  {Hradil}}]{Mogilevtsev2012}%
  \BibitemOpen
  \bibfield  {author} {\bibinfo {author} {\bibfnamefont {D.}~\bibnamefont
  {Mogilevtsev}}, \bibinfo {author} {\bibfnamefont {J.}~\bibnamefont {{\v
  R}eh{\'a}{\v c}ek}}, \ and\ \bibinfo {author} {\bibfnamefont
  {Z.}~\bibnamefont {Hradil}},\ }\href
  {http://stacks.iop.org/1367-2630/14/i=9/a=095001} {\bibfield  {journal}
  {\bibinfo  {journal} {New Journal of Physics},\ }\textbf {\bibinfo {volume}
  {14}},\ \bibinfo {pages} {095001} (\bibinfo {year} {2012})}\BibitemShut
  {NoStop}%
\bibitem [{\citenamefont {Teo}\ \emph {et~al.}(2012)\citenamefont {Teo},
  \citenamefont {Stoklasa}, \citenamefont {Englert}, \citenamefont {\ifmmode
  \check{R}\else \v{R}\fi{}eh\'a\ifmmode~\check{c}\else \v{c}\fi{}ek},\ and\
  \citenamefont {Hradil}}]{Teo2012}%
  \BibitemOpen
  \bibfield  {author} {\bibinfo {author} {\bibfnamefont {Y.~S.}\ \bibnamefont
  {Teo}}, \bibinfo {author} {\bibfnamefont {B.}~\bibnamefont {Stoklasa}},
  \bibinfo {author} {\bibfnamefont {B.-G.}\ \bibnamefont {Englert}}, \bibinfo
  {author} {\bibfnamefont {J.}~\bibnamefont {\ifmmode \check{R}\else
  \v{R}\fi{}eh\'a\ifmmode~\check{c}\else \v{c}\fi{}ek}}, \ and\ \bibinfo
  {author} {\bibfnamefont {Z.~c.~v.}\ \bibnamefont {Hradil}},\ }\Doi
  {10.1103/PhysRevA.85.042317} {\bibfield  {journal} {\bibinfo  {journal}
  {Phys. Rev. A},\ }\textbf {\bibinfo {volume} {85}},\ \bibinfo {pages}
  {042317} (\bibinfo {year} {2012})}\BibitemShut {NoStop}%
\bibitem [{\citenamefont {Merkel}\ \emph {et~al.}(2012)\citenamefont {Merkel},
  \citenamefont {Gambetta}, \citenamefont {Smolin}, \citenamefont {Poletto},
  \citenamefont {C{\'o}rcoles}, \citenamefont {Johnson},\ and\ \citenamefont
  {Colm A.~Ryan}}]{Merkel2012}%
  \BibitemOpen
  \bibfield  {author} {\bibinfo {author} {\bibfnamefont {S.~T.}\ \bibnamefont
  {Merkel}}, \bibinfo {author} {\bibfnamefont {J.~M.}\ \bibnamefont
  {Gambetta}}, \bibinfo {author} {\bibfnamefont {J.~A.}\ \bibnamefont
  {Smolin}}, \bibinfo {author} {\bibfnamefont {S.}~\bibnamefont {Poletto}},
  \bibinfo {author} {\bibfnamefont {A.~D.}\ \bibnamefont {C{\'o}rcoles}},
  \bibinfo {author} {\bibfnamefont {B.~R.}\ \bibnamefont {Johnson}}, \ and\
  \bibinfo {author} {\bibfnamefont {M.~S.}\ \bibnamefont {Colm A.~Ryan}},\
  }\href@noop {} {\bibfield  {journal} {\bibinfo  {journal} {arXiv:1211.0322
  [quant-ph]}} (\bibinfo {year} {2012})}\BibitemShut {NoStop}%
\bibitem [{\citenamefont {Altepeter}\ \emph {et~al.}(2005)\citenamefont
  {Altepeter}, \citenamefont {Jeffrey}, \citenamefont {Kwiat}, \citenamefont
  {Tanzilli}, \citenamefont {Gisin},\ and\ \citenamefont
  {Ac\'in}}]{Altepeter2005}%
  \BibitemOpen
  \bibfield  {author} {\bibinfo {author} {\bibfnamefont {J.~B.}\ \bibnamefont
  {Altepeter}}, \bibinfo {author} {\bibfnamefont {E.~R.}\ \bibnamefont
  {Jeffrey}}, \bibinfo {author} {\bibfnamefont {P.~G.}\ \bibnamefont {Kwiat}},
  \bibinfo {author} {\bibfnamefont {S.}~\bibnamefont {Tanzilli}}, \bibinfo
  {author} {\bibfnamefont {N.}~\bibnamefont {Gisin}}, \ and\ \bibinfo {author}
  {\bibfnamefont {A.}~\bibnamefont {Ac\'in}},\ }\Doi
  {10.1103/PhysRevLett.95.033601} {\bibfield  {journal} {\bibinfo  {journal}
  {Phys. Rev. Lett.},\ }\textbf {\bibinfo {volume} {95}},\ \bibinfo {pages}
  {033601} (\bibinfo {year} {2005})}\BibitemShut {NoStop}%
\bibitem [{\citenamefont {{\v R}eh{\'a}{\v c}ek}\ \emph
  {et~al.}(2007)\citenamefont {{\v R}eh{\'a}{\v c}ek}, \citenamefont {Hradil},
  \citenamefont {Knill},\ and\ \citenamefont {Lvovsky}}]{Rehacek2007}%
  \BibitemOpen
  \bibfield  {author} {\bibinfo {author} {\bibfnamefont {J.}~\bibnamefont {{\v
  R}eh{\'a}{\v c}ek}}, \bibinfo {author} {\bibfnamefont {Z.}~\bibnamefont
  {Hradil}}, \bibinfo {author} {\bibfnamefont {E.}~\bibnamefont {Knill}}, \
  and\ \bibinfo {author} {\bibfnamefont {A.~I.}\ \bibnamefont {Lvovsky}},\
  }\Doi {10.1103/PhysRevA.75.042108} {\bibfield  {journal} {\bibinfo  {journal}
  {Phys. Rev. A},\ }\textbf {\bibinfo {volume} {75}},\ \bibinfo {pages}
  {042108} (\bibinfo {year} {2007})}\BibitemShut {NoStop}%
\bibitem [{\citenamefont {Adamson}\ and\ \citenamefont
  {Steinberg}(2010)}]{Adamson2010}%
  \BibitemOpen
  \bibfield  {author} {\bibinfo {author} {\bibfnamefont {R.~B.~A.}\
  \bibnamefont {Adamson}}\ and\ \bibinfo {author} {\bibfnamefont {A.~M.}\
  \bibnamefont {Steinberg}},\ }\Doi {10.1103/PhysRevLett.105.030406} {\bibfield
   {journal} {\bibinfo  {journal} {Phys. Rev. Lett.},\ }\textbf {\bibinfo
  {volume} {105}},\ \bibinfo {pages} {030406} (\bibinfo {year}
  {2010})}\BibitemShut {NoStop}%
\bibitem [{\citenamefont {Bogdanov}\ \emph {et~al.}(2010)\citenamefont
  {Bogdanov}, \citenamefont {Brida}, \citenamefont {Genovese}, \citenamefont
  {Kulik}, \citenamefont {Moreva},\ and\ \citenamefont
  {Shurupov}}]{Bogdanov2010}%
  \BibitemOpen
  \bibfield  {author} {\bibinfo {author} {\bibfnamefont {Y.~I.}\ \bibnamefont
  {Bogdanov}}, \bibinfo {author} {\bibfnamefont {G.}~\bibnamefont {Brida}},
  \bibinfo {author} {\bibfnamefont {M.}~\bibnamefont {Genovese}}, \bibinfo
  {author} {\bibfnamefont {S.~P.}\ \bibnamefont {Kulik}}, \bibinfo {author}
  {\bibfnamefont {E.~V.}\ \bibnamefont {Moreva}}, \ and\ \bibinfo {author}
  {\bibfnamefont {A.~P.}\ \bibnamefont {Shurupov}},\ }\Doi
  {10.1103/PhysRevLett.105.010404} {\bibfield  {journal} {\bibinfo  {journal}
  {Phys. Rev. Lett.},\ }\textbf {\bibinfo {volume} {105}},\ \bibinfo {pages}
  {010404} (\bibinfo {year} {2010})}\BibitemShut {NoStop}%
\bibitem [{\citenamefont {Yamagata}(2011)}]{Yamagata2011}%
  \BibitemOpen
  \bibfield  {author} {\bibinfo {author} {\bibfnamefont {K.}~\bibnamefont
  {Yamagata}},\ }\href@noop {} {\bibfield  {journal} {\bibinfo  {journal}
  {International Journal of Quantum Information},\ }\textbf {\bibinfo {volume}
  {9}},\ \bibinfo {pages} {1167} (\bibinfo {year} {2011})}\BibitemShut
  {NoStop}%
\bibitem [{\citenamefont {Brida}\ \emph {et~al.}(2011)\citenamefont {Brida},
  \citenamefont {Degiovanni}, \citenamefont {Florio}, \citenamefont {Genovese},
  \citenamefont {Giorda}, \citenamefont {Meda}, \citenamefont {Paris},\ and\
  \citenamefont {Shurupov}}]{Brida2011}%
  \BibitemOpen
  \bibfield  {author} {\bibinfo {author} {\bibfnamefont {G.}~\bibnamefont
  {Brida}}, \bibinfo {author} {\bibfnamefont {I.~P.}\ \bibnamefont
  {Degiovanni}}, \bibinfo {author} {\bibfnamefont {A.}~\bibnamefont {Florio}},
  \bibinfo {author} {\bibfnamefont {M.}~\bibnamefont {Genovese}}, \bibinfo
  {author} {\bibfnamefont {P.}~\bibnamefont {Giorda}}, \bibinfo {author}
  {\bibfnamefont {A.}~\bibnamefont {Meda}}, \bibinfo {author} {\bibfnamefont
  {M.~G.~A.}\ \bibnamefont {Paris}}, \ and\ \bibinfo {author} {\bibfnamefont
  {A.~P.}\ \bibnamefont {Shurupov}},\ }\Doi {10.1103/PhysRevA.83.052301}
  {\bibfield  {journal} {\bibinfo  {journal} {Phys. Rev. A},\ }\textbf
  {\bibinfo {volume} {83}},\ \bibinfo {pages} {052301} (\bibinfo {year}
  {2011})}\BibitemShut {NoStop}%
\bibitem [{\citenamefont {Bogdanov}\ \emph {et~al.}(2011)\citenamefont
  {Bogdanov}, \citenamefont {Brida}, \citenamefont {Bukeev}, \citenamefont
  {Genovese}, \citenamefont {Kravtsov}, \citenamefont {Kulik}, \citenamefont
  {Moreva}, \citenamefont {Soloviev},\ and\ \citenamefont
  {Shurupov}}]{Bogdanov2011}%
  \BibitemOpen
  \bibfield  {author} {\bibinfo {author} {\bibfnamefont {Y.~I.}\ \bibnamefont
  {Bogdanov}}, \bibinfo {author} {\bibfnamefont {G.}~\bibnamefont {Brida}},
  \bibinfo {author} {\bibfnamefont {I.~D.}\ \bibnamefont {Bukeev}}, \bibinfo
  {author} {\bibfnamefont {M.}~\bibnamefont {Genovese}}, \bibinfo {author}
  {\bibfnamefont {K.~S.}\ \bibnamefont {Kravtsov}}, \bibinfo {author}
  {\bibfnamefont {S.~P.}\ \bibnamefont {Kulik}}, \bibinfo {author}
  {\bibfnamefont {E.~V.}\ \bibnamefont {Moreva}}, \bibinfo {author}
  {\bibfnamefont {A.~A.}\ \bibnamefont {Soloviev}}, \ and\ \bibinfo {author}
  {\bibfnamefont {A.~P.}\ \bibnamefont {Shurupov}},\ }\Doi
  {10.1103/PhysRevA.84.042108} {\bibfield  {journal} {\bibinfo  {journal}
  {Phys. Rev. A},\ }\textbf {\bibinfo {volume} {84}},\ \bibinfo {pages}
  {042108} (\bibinfo {year} {2011})}\BibitemShut {NoStop}%
\bibitem [{\citenamefont {Kalev}\ \emph
  {et~al.}(2012){\natexlab{b}}\citenamefont {Kalev}, \citenamefont {Shang},\
  and\ \citenamefont {Englert}}]{Kalev2012}%
  \BibitemOpen
  \bibfield  {author} {\bibinfo {author} {\bibfnamefont {A.}~\bibnamefont
  {Kalev}}, \bibinfo {author} {\bibfnamefont {J.}~\bibnamefont {Shang}}, \ and\
  \bibinfo {author} {\bibfnamefont {B.-G.}\ \bibnamefont {Englert}},\ }\Doi
  {10.1103/PhysRevA.85.052115} {\bibfield  {journal} {\bibinfo  {journal}
  {Phys. Rev. A},\ }\textbf {\bibinfo {volume} {85}},\ \bibinfo {pages}
  {052115} (\bibinfo {year} {2012}{\natexlab{b}})}\BibitemShut {NoStop}%
\bibitem [{\citenamefont {Flueraru}\ \emph {et~al.}(2008)\citenamefont
  {Flueraru}, \citenamefont {Latoui}, \citenamefont {Besse},\ and\
  \citenamefont {Legendre}}]{Flueraru2008}%
  \BibitemOpen
  \bibfield  {author} {\bibinfo {author} {\bibfnamefont {C.}~\bibnamefont
  {Flueraru}}, \bibinfo {author} {\bibfnamefont {S.}~\bibnamefont {Latoui}},
  \bibinfo {author} {\bibfnamefont {J.}~\bibnamefont {Besse}}, \ and\ \bibinfo
  {author} {\bibfnamefont {P.}~\bibnamefont {Legendre}},\ }\Doi
  {10.1109/TIM.2007.913752} {\bibfield  {journal} {\bibinfo  {journal}
  {Instrumentation and Measurement, IEEE Transactions on},\ }\textbf {\bibinfo
  {volume} {57}},\ \bibinfo {pages} {731 } (\bibinfo {year} {2008})},\ ISSN
  \bibinfo {issn} {0018-9456}\BibitemShut {NoStop}%
\bibitem [{\citenamefont {Wasilewski}\ \emph {et~al.}(2006)\citenamefont
  {Wasilewski}, \citenamefont {Wasylczyk}, \citenamefont {Kolenderski},
  \citenamefont {Banaszek},\ and\ \citenamefont {Radzewicz}}]{Wasilewski2006a}%
  \BibitemOpen
  \bibfield  {author} {\bibinfo {author} {\bibfnamefont {W.}~\bibnamefont
  {Wasilewski}}, \bibinfo {author} {\bibfnamefont {P.}~\bibnamefont
  {Wasylczyk}}, \bibinfo {author} {\bibfnamefont {P.}~\bibnamefont
  {Kolenderski}}, \bibinfo {author} {\bibfnamefont {K.}~\bibnamefont
  {Banaszek}}, \ and\ \bibinfo {author} {\bibfnamefont {C.}~\bibnamefont
  {Radzewicz}},\ }\Doi {10.1364/OL.31.001130} {\bibfield  {journal} {\bibinfo
  {journal} {Opt. Lett.},\ }\textbf {\bibinfo {volume} {31}},\ \bibinfo {pages}
  {1130} (\bibinfo {year} {2006})}\BibitemShut {NoStop}%
\bibitem [{\citenamefont {Hradil}(1997)}]{Hradil1997}%
  \BibitemOpen
  \bibfield  {author} {\bibinfo {author} {\bibfnamefont {Z.}~\bibnamefont
  {Hradil}},\ }\Doi {10.1103/PhysRevA.55.R1561} {\bibfield  {journal} {\bibinfo
   {journal} {Phys. Rev. A},\ }\textbf {\bibinfo {volume} {55}},\ \bibinfo
  {pages} {R1561} (\bibinfo {year} {1997})}\BibitemShut {NoStop}%
\bibitem [{\citenamefont {Kaznady}\ and\ \citenamefont
  {James}(2009)}]{Kaznady2009}%
  \BibitemOpen
  \bibfield  {author} {\bibinfo {author} {\bibfnamefont {M.~S.}\ \bibnamefont
  {Kaznady}}\ and\ \bibinfo {author} {\bibfnamefont {D.~F.~V.}\ \bibnamefont
  {James}},\ }\Doi {10.1103/PhysRevA.79.022109} {\bibfield  {journal} {\bibinfo
   {journal} {Phys. Rev. A},\ }\textbf {\bibinfo {volume} {79}},\ \bibinfo
  {pages} {022109} (\bibinfo {year} {2009})}\BibitemShut {NoStop}%
\bibitem [{\citenamefont
  {Blume-Kohout}(2010){\natexlab{a}}}]{Blume-Kohout2010}%
  \BibitemOpen
  \bibfield  {author} {\bibinfo {author} {\bibfnamefont {R.}~\bibnamefont
  {Blume-Kohout}},\ }\href {http://stacks.iop.org/1367-2630/12/i=4/a=043034}
  {\bibfield  {journal} {\bibinfo  {journal} {New Journal of Physics},\
  }\textbf {\bibinfo {volume} {12}},\ \bibinfo {pages} {043034} (\bibinfo
  {year} {2010}{\natexlab{a}})}\BibitemShut {NoStop}%
\bibitem [{\citenamefont {Gross}\ \emph {et~al.}(2010)\citenamefont {Gross},
  \citenamefont {Liu}, \citenamefont {Flammia}, \citenamefont {Becker},\ and\
  \citenamefont {Eisert}}]{Gross2010a}%
  \BibitemOpen
  \bibfield  {author} {\bibinfo {author} {\bibfnamefont {D.}~\bibnamefont
  {Gross}}, \bibinfo {author} {\bibfnamefont {Y.-K.}\ \bibnamefont {Liu}},
  \bibinfo {author} {\bibfnamefont {S.~T.}\ \bibnamefont {Flammia}}, \bibinfo
  {author} {\bibfnamefont {S.}~\bibnamefont {Becker}}, \ and\ \bibinfo {author}
  {\bibfnamefont {J.}~\bibnamefont {Eisert}},\ }\Doi
  {10.1103/PhysRevLett.105.150401} {\bibfield  {journal} {\bibinfo  {journal}
  {Phys. Rev. Lett.},\ }\textbf {\bibinfo {volume} {105}},\ \bibinfo {pages}
  {150401} (\bibinfo {year} {2010})}\BibitemShut {NoStop}%
\bibitem [{\citenamefont {Teo}\ \emph {et~al.}(2011)\citenamefont {Teo},
  \citenamefont {Zhu}, \citenamefont {Englert}, \citenamefont {\ifmmode
  \check{R}\else \v{R}\fi{}eh\'a\ifmmode~\check{c}\else \v{c}\fi{}ek},\ and\
  \citenamefont {Hradil}}]{Teo2011}%
  \BibitemOpen
  \bibfield  {author} {\bibinfo {author} {\bibfnamefont {Y.~S.}\ \bibnamefont
  {Teo}}, \bibinfo {author} {\bibfnamefont {H.}~\bibnamefont {Zhu}}, \bibinfo
  {author} {\bibfnamefont {B.-G.}\ \bibnamefont {Englert}}, \bibinfo {author}
  {\bibfnamefont {J.}~\bibnamefont {\ifmmode \check{R}\else
  \v{R}\fi{}eh\'a\ifmmode~\check{c}\else \v{c}\fi{}ek}}, \ and\ \bibinfo
  {author} {\bibfnamefont {Z.~c.~v.}\ \bibnamefont {Hradil}},\ }\Doi
  {10.1103/PhysRevLett.107.020404} {\bibfield  {journal} {\bibinfo  {journal}
  {Phys. Rev. Lett.},\ }\textbf {\bibinfo {volume} {107}},\ \bibinfo {pages}
  {020404} (\bibinfo {year} {2011})}\BibitemShut {NoStop}%
\bibitem [{\citenamefont
  {Blume-Kohout}(2010){\natexlab{b}}}]{Blume-Kohout2010a}%
  \BibitemOpen
  \bibfield  {author} {\bibinfo {author} {\bibfnamefont {R.}~\bibnamefont
  {Blume-Kohout}},\ }\Doi {10.1103/PhysRevLett.105.200504} {\bibfield
  {journal} {\bibinfo  {journal} {Phys. Rev. Lett.},\ }\textbf {\bibinfo
  {volume} {105}},\ \bibinfo {pages} {200504} (\bibinfo {year}
  {2010}{\natexlab{b}})}\BibitemShut {NoStop}%
\bibitem [{\citenamefont {Ng}\ and\ \citenamefont {Englert}(2012)}]{Ng2012}%
  \BibitemOpen
  \bibfield  {author} {\bibinfo {author} {\bibfnamefont {H.~K.}\ \bibnamefont
  {Ng}}\ and\ \bibinfo {author} {\bibfnamefont {B.-G.}\ \bibnamefont
  {Englert}},\ }\href@noop {} {\bibfield  {journal} {\bibinfo  {journal}
  {arXiv:1202.5136 [quant-ph]}} (\bibinfo {year} {2012})}\BibitemShut {NoStop}%
\bibitem [{\citenamefont {Christandl}\ and\ \citenamefont
  {Renner}(2011)}]{Christandl2011}%
  \BibitemOpen
  \bibfield  {author} {\bibinfo {author} {\bibfnamefont {M.}~\bibnamefont
  {Christandl}}\ and\ \bibinfo {author} {\bibfnamefont {R.}~\bibnamefont
  {Renner}},\ }\href@noop {} {\bibfield  {journal} {\bibinfo  {journal}
  {arXiv:1108.5329v1 [quant-ph]}} (\bibinfo {year} {2011})}\BibitemShut
  {NoStop}%
\bibitem [{\citenamefont {Blume-Kohout}(2012)}]{Blume-Kohout2012}%
  \BibitemOpen
  \bibfield  {author} {\bibinfo {author} {\bibfnamefont {R.}~\bibnamefont
  {Blume-Kohout}},\ }\href@noop {} {\bibfield  {journal} {\bibinfo  {journal}
  {arXiv:1202.5270v1 [quant-ph]}} (\bibinfo {year} {2012})}\BibitemShut
  {NoStop}%
\bibitem [{\citenamefont {Sabatke}\ \emph {et~al.}(2000)\citenamefont
  {Sabatke}, \citenamefont {Descour}, \citenamefont {Dereniak}, \citenamefont
  {Sweatt}, \citenamefont {Kemme},\ and\ \citenamefont {Phipps}}]{Sabatke2000}%
  \BibitemOpen
  \bibfield  {author} {\bibinfo {author} {\bibfnamefont {D.~S.}\ \bibnamefont
  {Sabatke}}, \bibinfo {author} {\bibfnamefont {M.~R.}\ \bibnamefont
  {Descour}}, \bibinfo {author} {\bibfnamefont {E.~L.}\ \bibnamefont
  {Dereniak}}, \bibinfo {author} {\bibfnamefont {W.~C.}\ \bibnamefont
  {Sweatt}}, \bibinfo {author} {\bibfnamefont {S.~A.}\ \bibnamefont {Kemme}}, \
  and\ \bibinfo {author} {\bibfnamefont {G.~S.}\ \bibnamefont {Phipps}},\ }\Doi
  {10.1364/OL.25.000802} {\bibfield  {journal} {\bibinfo  {journal} {Opt.
  Lett.},\ }\textbf {\bibinfo {volume} {25}},\ \bibinfo {pages} {802} (\bibinfo
  {year} {2000})}\BibitemShut {NoStop}%
\end{thebibliography}
\end{document}